\documentclass{article}
\usepackage{smc2020}
\usepackage{times}
\usepackage{ifpdf}
\usepackage{subfigure}
\usepackage[english]{babel}
\usepackage{cite}
\usepackage[pdftex]{graphicx} 
\usepackage{epstopdf}

\def\papertitle{Exploring Inherent Properties of the Monophonic Melody of Songs}
\def\firstauthor{First author}
\def\secondauthor{Second author}
\def\thirdauthor{Third author}

\newif\ifpdf
\ifx\pdfoutput\relax
\else
\ifcase\pdfoutput
\pdffalse
\else
\pdftrue
\fi

\ifpdf 
\usepackage[pdftex,
pdftitle={\papertitle},
pdfauthor={\firstauthor, \secondauthor, \thirdauthor},
bookmarksnumbered, 
pdfstartview=XYZ 
]{hyperref}

\usepackage[pdftex]{graphicx}
\DeclareGraphicsExtensions{.pdf,.jpeg,.png}

\usepackage[figure,table]{hypcap}

\else 
\usepackage[dvips,
bookmarksnumbered, 
pdfstartview=XYZ 
]{hyperref}  

\usepackage[dvips]{epsfig,graphicx}
\graphicspath{{./figures/}}
\DeclareGraphicsExtensions{.eps}

\usepackage[figure,table]{hypcap}
\fi

\hypersetup{
	colorlinks,%
	citecolor=black,%
	filecolor=black,%
	linkcolor=black,%
	urlcolor=black
}

\title{\papertitle}

\threeauthors
{Zehao Wang} {Peking University \\ %
	{\tt \href{mailto:water45wzh@pku.edu.cn}{water45wzh@pku.edu.cn}}}
{Shicheng Zhang} {University of Illinois at Urbana-Champaign	 \\ %
	{\tt \href{mailto:sz18@illinois.edu}{sz18@illinois.edu}}}
{Xiaoou Chen} {Peking University\\ %
	{\tt \href{mailto:chenxiaoou@pku.edu.cn}{chenxiaoou@pku.edu.cn}}}

\begin{document}
	\capstartfalse
	\maketitle
	\capstarttrue
	\begin{abstract}
		Melody is one of the most important components in music. Unlike other components in music theory, such as harmony and counterpoint, computable features for melody is urgently in need.  These features are highly demanded as data-driven methods dominating the fields such as musical information retrieval and automatic music composition. To boost the performance of deep-learning-related musical tasks, we propose a set of interpretable features on monophonic melody for computational purposes. These features are defined not only in mathematical form, but also with some considerations on composers’ intuition. For example, the Melodic Center of Gravity can reflect the sentence-wise contour of the melody, the local /global melody dynamics quantifies the dynamics of a melody that couples pitch and time in a sentence. We found that these features are considered by people universally in many genres of songs, even for atonal composition practices. Hopefully, these melodic features can provide novel inspiration for future researchers as a tool in the field of MIR and automatic composition.
	\end{abstract}

	\section{Introduction}\label{sec:introduction}
	In recent years, with the improvement of technology, especially machine learning and deep learning technology, the music industry has undergone tremendous changes, and interdisciplinary research fields such as automatic music generation, music recommendation systems, and automatic music evaluation have gradually developed. All most all the current state-of-the-art methods in these fields are closely related to deep learning. 
	
	As we know, in modern deep learning technology, the super large amount of high-dimensional data and effective feature extraction are the keys to success. However, unlike the image, audio or natural language, the popular deep learning technology is not easy to perform feature engineering on very low-dimensional music information, especially the monophonic melody. Here, we only discuss the music of Midi / Sheet Music or similar format, without considering the audio after recording and mixing. The main difficulties are the following. 
	\begin{itemize}
		\item First, it is difficult for music to obtain a very large dataset or corpus with a uniform format and complete labeling, like image and natural language, and it is necessary to have professionally trained personnel to organize such a music dataset.
		\item Second, unlike audio, music data (for example in Midi format) is very low-dimensional, and multi-track music always can be highly structured, existing deep learning technologies does not perform well on such data. 
		\item Third, for image, some mathematicians have used the methods of partial differential equations and inverse problems to establish a set of rigorous theories; for audio, digital signal processing provides a number of interpretable feature extraction methods; for natural language, there is computational linguistics based on linguistics which provides an interpretable research method.
	\end{itemize}
	
	In some branches of music theory, such as Harmony, Counterpoint and Orchestration, there are some self-consistent theories for music creation and evaluation, and even theoretical explanations based on mathematics, physics, etc. Except for theoretical explanations based on audio or acoustics, for music data based on Sheet Music or Midi format, some of the existing methods use more abstract mathematical theories, such as Abstract Algebra and Topology, to try to explain some of the underlying phenomena. Their common problem is that they are mathematically beautiful, but because they take high computational complexity or cannot be computed by machine at all, we cannot use it to help us design algorithms. Other existing methods are more like statistics of midi events, such as counting their histograms, without touching composition theory or human perception (whether in the perspective of creation or appreciation). These low-level semantic features are rigid and difficult to use flexibly. For example, if we directly intervene in the automatic generation algorithm with a given melodic interval sequence and rhythm sequence as conditions, this is rigid and inconsistent with the way humans compose music, and maybe the melody will be, and even the generated melody will be a mess. But if we don't put any conditions on the generation algorithm, they will become uncontrollable, and even the generated melody will still be a mess.
	
	Therefore, for the music data we are discussing here, there is still a lack of a set of high/intermediate-level semantic, human-perception related, interpretable and computable feature extraction methods, such as Melspectrogram in audio, to enable existing deep learning-based algorithms for automatic music generation or automatic music evaluation more interpretable or controllable, or they can be used to design corresponding rule-based algorithms.
	
	In this paper, we focus on the monophonic melody of songs, explore its inherent properties. This means that, we only study the monophonic melody, no matter which temperaments, scale modes, even whether tonal or atonal they are, and we do not consider the relationship between melody \& harmony or melody \& lyrics, only focus on the inherent properties of the monophonic melody of songs. Inspired from Psychology and Musicology/Composition Theory, we propose some coarse-grained features of melody mathematically from a macro perspective, like
	\begin{itemize}
		\item  \textbf{Melodic Center of Gravity/Melody Center:} Used to demonstrate the sentence-wise contour of the melody flow.
		\item \textbf{Local/Global Melody Dynamic:} Used to study the dynamic inside a sentence (mainly about the melody which couples the pitch series and rhythm).
		\item \textbf{Rhythm Dynamic:} Used to study the dynamic inside a sentence (mainly about the rhythm).
	\end{itemize}
	
	All the above features are designed based on human perception and human composition process, and all have their specific musical significance. The framework we propose is flexible, and some of these modules can be adjusted or other quantities constructed in accordance with the assumptions in the framework to solve different problems. To the best of our knowledge, this is the first work to study the inherent properties of the melody in this novel and rigorous way. Since images, audio, and natural languages have all experienced interpretable feature extraction research based on human perception before entering the era of deep learning, they also designed the corresponding most advanced deep learning technologies based on the work of these predecessors. The lack of similar work in music may be the biggest reason for the poor performance of existing deep learning methods. Therefore, we think that similar work needs to be done on music, so this article may be one of the pioneers.
	
	\section{Background and Preliminaries}
	First of all, we must deeply understand how people write melody, what people think about when writing a melody, and the general characteristics of melody written by pros. 
	
	\subsection{How do people write melody?}
	When pros write melody, they always start with the \textbf{Motif}\footnote{\textit{In music, a motif About this sound(pronunciation) (also motive) is a short musical phrase, a salient recurring figure, musical fragment or succession of notes that has some special importance in or is characteristic of a composition: "The motive is the smallest structural unit possessing thematic identity".} ----From Wikipedia}--the basic module in a song/music. Then they use some composition techniques, which they may call transformation\footnote{such as translation, rotation, and scaling, sometimes they may use modulation and tonicization}, to make a song logically and not boring from their basic motifs. But these are not enough. In order to write a complete song, we also need to consider about the structure, which is called \textbf{Musical Form}\footnote{In pop music, people always use \textit{verse, pre-chorus, chorus} as the notations of structure. In classical music, we have some concept of forms such as \textit{sonota, rondo}.  In Chinese, we also use \textit{Qi Cheng Zhuan He} to denote similar meanings.}. 
	
\begin{figure*}
		\subfigure{
			\begin{minipage}[]{0.7\linewidth}
				\includegraphics[height=9cm]{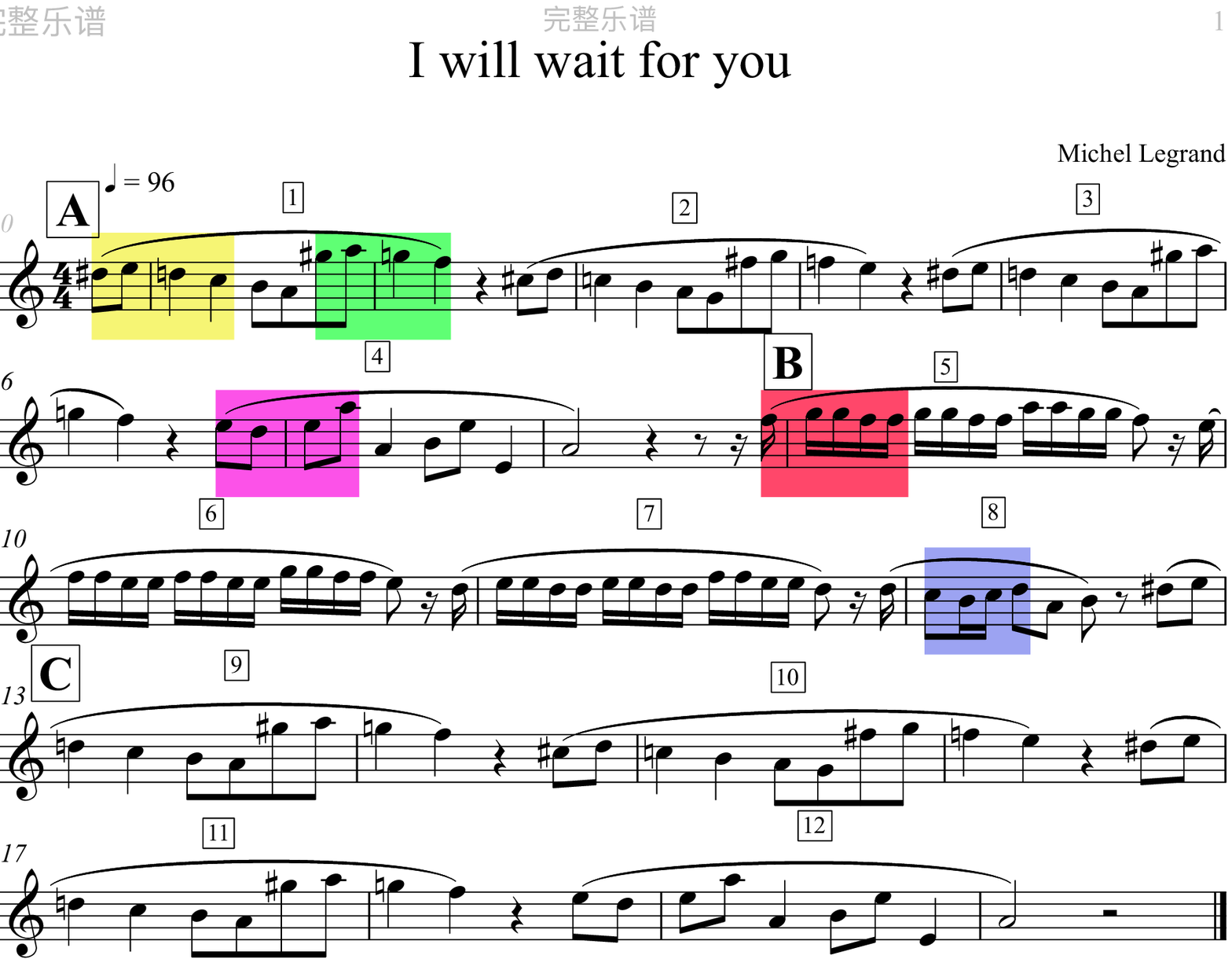}
			\end{minipage}
			\begin{minipage}[]{0.2\linewidth}
				\includegraphics[scale=0.30]{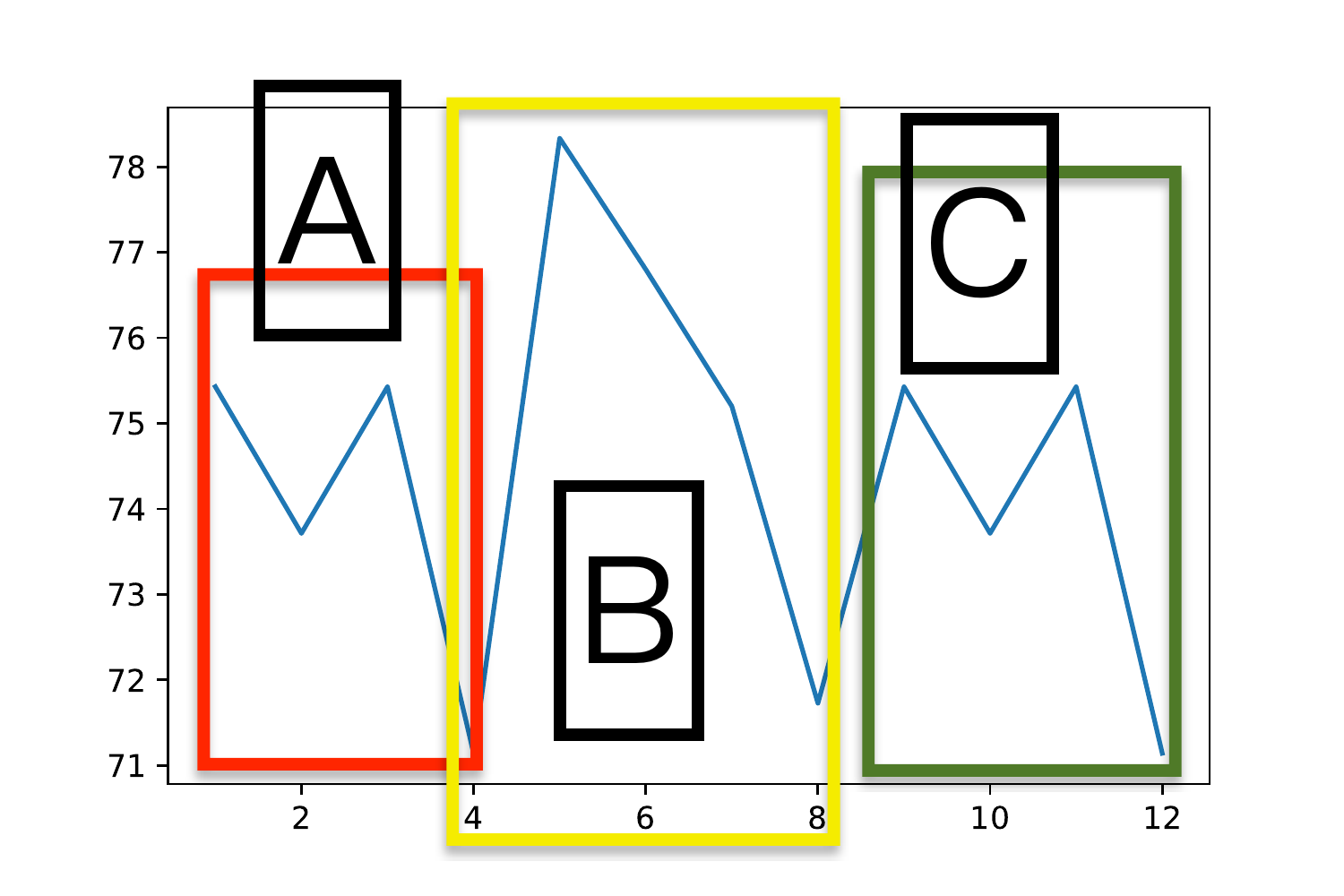}
				\includegraphics[scale=0.30]{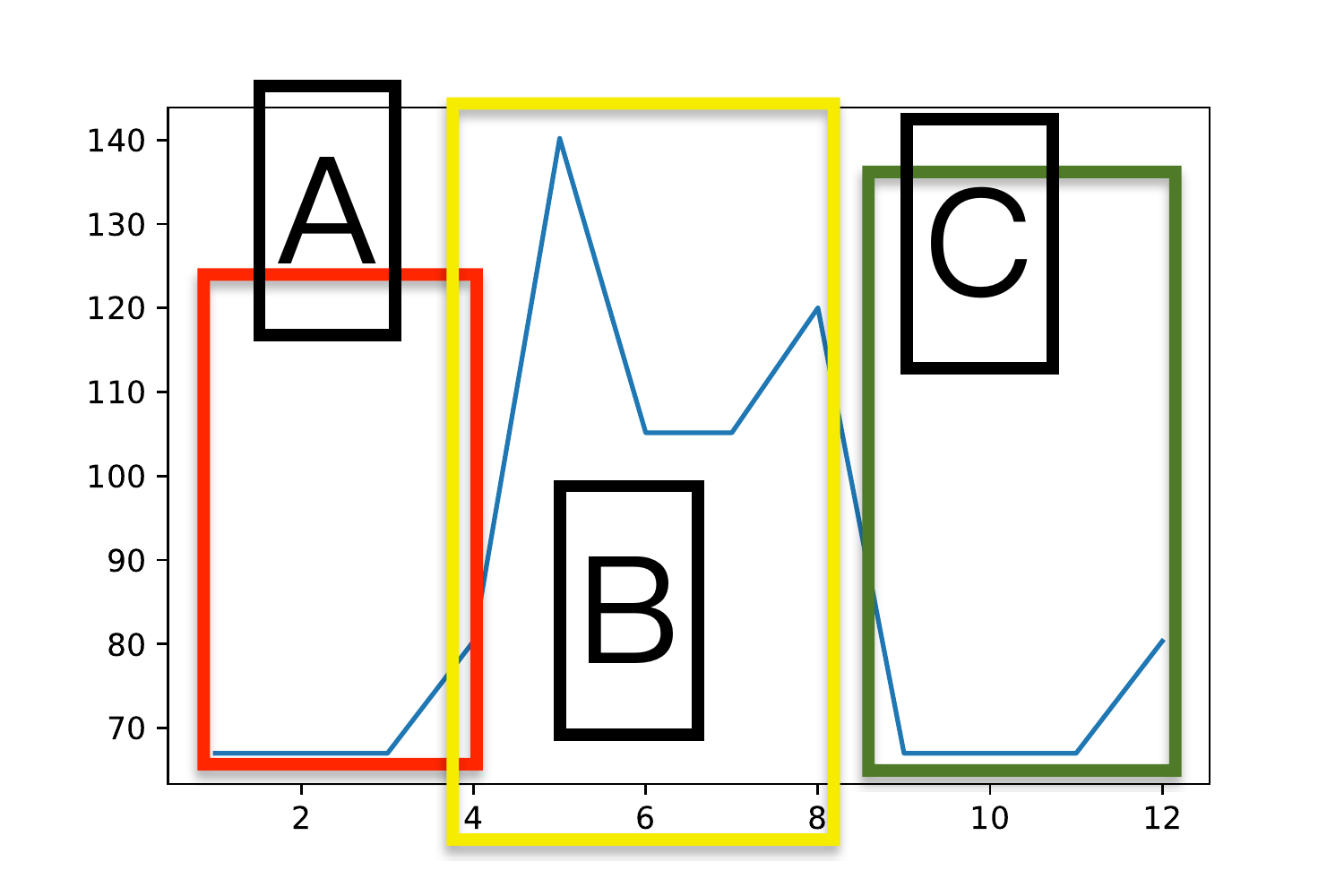}
				\includegraphics[scale=0.30]{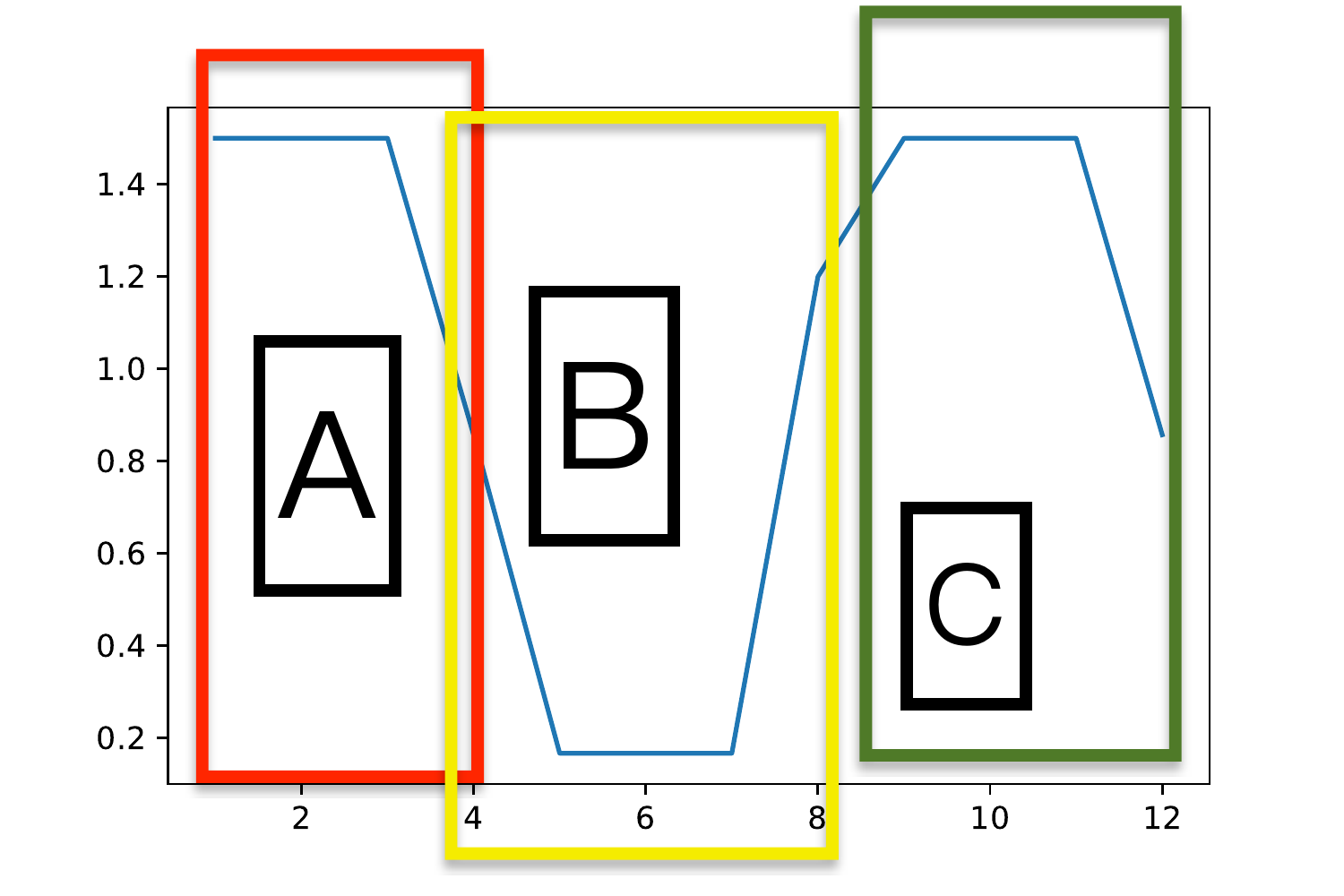}
			\end{minipage}%
		}%
		\caption{Brief analysis of this song: Motif: The yellow box is the motif for this song, the green one is a translation of the motif, the pink one is an inverse(rotation) of it, the red one is scaling of the motif, and the purple one is inverse(rotation)+scaling of it. Musical Form: A denotes the verse, B denotes the chorus, and C denotes the Recapitulation. Number above the slur means the index of this sentence. Three figures in the second column denote sentence-wise MCG, GMD and RD respectively. Their x-axis is the index of sentences. And we choose non-normalized L$\infty$-GMD for this song (choosing non-normalized is because the sentences in this song are neat). }
		\label{cherbourg}
	\end{figure*}

	Here we take a famous song, \textit{I will wait for you} by \textit{Michel Legrand}\footnote{\href{http://michellegrandofficial.com/}{http://michellegrandofficial.com/}}, as an example to demonstrate these concepts mentioned above. It is shown in Fig. \ref{cherbourg}. All the analyses is in the caption of Fig. \ref{cherbourg}.
	
	From the example above, we can see/hear that the composer will design the undulations between different parts, in other words the overall melody contour; then within each part, the sentences are also related each other; the dynamic/tension contrast of the sentences between different parts is distinct.
	
	\subsection{General Characteristics of Good Melodies}
	Regardless of composer's personal characteristics and different genres/styles, here we only talk about the general characteristics of good melodies.
	
	First of all, \textbf{a good melody should be easy and comfortable to sing}. In other words, it should have smooth rhythm and melody line, match the range of human voice, and last but not least, it should have no ridiculous oversized jumps.
	
	Then, \textbf{a good melody should be melodious}. It means the melody should match the chord of the corresponding accompany, and it has catchy chord progression\footnote{like chord progression of \textit{Canon in D} by \textit{Pachelbel}, or IV-V-III-VI-II-V-I which is very popular in C-pop and old J-pop.}. In addition to the relationship between melody and chord, it should also have reasonable melody dynamic in different parts of it, such as the contrast between verse and chorus.
	
	\textbf{Some of the famous melodies written by pros always have unexpected surprises}, such as reharmonization in their own way, or personal melody lines. These characteristics are why we can easily distinguish them from thousands of songs, or we can say why these pros are called pros and they are famous.
	
	\subsection{Related Works}
	As we said above, there are some explanatory works on music theory, some of them such as \cite{tymoczko2010geometry}\cite{mazzola2012topos} are explained based on the theory of mathematics or physics, theories they used are abstract and difficult to understand and calculate, or even can not be computed by computer. 
	
	Some studies, such as \cite{mckay2006jsymbolic}\cite{mckay2018jsymbolic}, directly perform some statistics on midi events (such as histograms of pitch and duration), and then use these statistical quantities as features for machine learning algorithms to use. But the limitation of this method is too low-level semantics, far from human perception, and even their feature dimensions are higher than midi or symbolic music itself,  so it is difficult to perform well for deep learning algorithms in some tasks such as automatic music generation and objective evaluation.
	
	There is also some work from a psychological perspective, such as \cite{quinto2013emotional}\cite{thomassen1982melodic}\cite{condit2019deconstructing}\cite{toussaint2012pairwise}\cite{temperley2004cognition}. They have proposed some useful quantities like nPVI to measure the dynamic of rhythm, and used these quantities as tools to study the rhythm relationship between speech and music, also proved that these quantities are consistent with human perception  from the perspective of psychological experiments. However, their common problem is that they ignore the logical considerations of human composition process. In other words, they ignore the inherent logical structure of music and do not propose a rigorous framework for music research.
	
	Although some studies on music using very abstract mathematical theories may seem to clarify its internal structure, they can hardly be calculated, so that it is hardly to be used for algorithm design.
	
	Some existing music generation algorithms using deep learning technology, such as \cite{huang2018music}\cite{huang2020pop}\cite{yang2017midinet}\cite{dong2018musegan}, some of them which seem to use advanced technology, are controlled directly using a fixed harmony progression or melody interval relationship, but their generated music is neither illogical, unmelodious, or even unstructured. These rigid and low-level semantic features do not fit well with human perception and composition. As we said before, it enlightens us to start with the human composition process, design some feature extraction methods that are consistent with human perception, and control the automatic music generation process with these coarse-grained features.

	\section{Study Melody in a Rigorous Mathematical Way}
	In this section, we will define some quantities mathematically with perceptual and musical explanations from different aspects of the melody, and give some calculation and visualization examples in \ref{example}.
	
	First we give a mathematical definition for note, rest and melody.
	
	\hspace*{\fill}
	
	\noindent \textbf{Definition.} Given a \textbf{\textit{Note}} $N=(P,d)$, it has two basic components: (relative) pitch $P$ and (relative) duration $d$. A \textbf{\textit{Rest}} $E$ only has duration components, so we use $d_E$ to represent its duration. A monophonic \textbf{\textit{Melody}} is defined as $M=K_1K_2...K_n$ where $K$ is a note $N$ or a rest $E$.
	
	\hspace*{\fill}
	
	Here we only consider pitch and duration aspects of the note, and note series of the melody, which means we do not consider the bar. And notice that we write relative pitch and relative duration above, it means in some features we need to choose the units of pitch and duration, we will discuss it later.
	
	\subsection{Sentence-wise Contour of Melody}
	When people listen to a song, it is easy to feel whether the tone of the sentence is high or low, or that they can easily feel which of the two sentences has a higher tone (the tone here represents the macro perception of all the pitches of a whole sentence). But in a sentence, there may be a lot of fragmented notes, and even they may produce poly-words on one note or poly-note on one word due to the problem of lyrics language. Even in such complex sentences, people can still perceive the sentence-wise movement process. So we need some tools to help machines get the quantitative measurements of this phenomena. For this purpose, here we will introduce the \textbf{Melody Center of Gravity}.
	
	\hspace*{\fill}
	
	\noindent \textbf{Definition.} Given a monophonic melody without rest, we formalize it as $M=N_1N_2...N_n$, where $N_i=(P_i.d_i)$, and use $|M|$ to denote its length $n$. Then we define $MCG_M$, the \textbf{\textit{Melodic Center of Gravity (MCG)}} of this given melody $M$ as follows,
	\
	\begin{equation}
	MCG_M=\frac{\sum_{i=1}^{n}d_iP_i}{\sum_{i=1}^{n}d_i}.
	\end{equation}
	
	Due to the homogeneity of $d_i$ in the above formula, we don not need to consider its unit. 
	
	As for the relative position of the pitch, we may adjust the weights of the internal and external tones (both for harmony and scale), not simply use semitone as the basic unit. 
	
	Note that this definition is only about melodies without rest. Considering the sense of listening to the whole sentence and the expression way of people playing or singing, if the rest is in a musical sentence, we can ignore it and add its duration to the previous note during the calculation of this sentence.
	
	Although we have been emphasizing the sentence-wise MCG calculation here, we can also calculate MCG of each bar or full song to measure the corresponding perceptual features.
	
	Remark: If we set all durations to be 1, then we have a special type of MCG, which is called \textbf{Melody Center}, in some literature (\cite{mckay2006jsymbolic}\cite{mckay2018jsymbolic}) they also call it \textbf{Mean Pitch}. 
	
	\begin{equation}
	MG_M=\frac{\sum_{i=1}^{n}P_i}{n}
	\end{equation}
	
\begin{figure}
		\centering
		\includegraphics[scale=0.7]{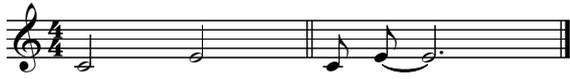}
		\caption{Example for MCG and MC}
		\label{mcgex}
	\end{figure}

	However, we do not recommend this. We can see the following example in Fig. \ref{mcgex}. MCs of these two melody are the same, but it ’s clear that the latter is higher overall. So from this, we can see the need to propose MCG which is more flexible and couples pitch and time\footnote{In practice, we can choose proper pitch unit and duration unit to change to adjust the weight of different notes according to factors such as different scales or chords.}.
	
	\subsection{Dynamic inside a Sentence}
\begin{figure}
		\centering
		\includegraphics[scale=0.7]{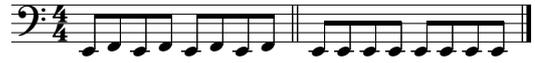}
		\caption{Bass melody line of \textit{Shark Theme} from "Jaws" by \textit{John Williams}}
		\label{jaw}
	\end{figure}

	Let us begin with a famous film music -- \textit{Shark Theme} from "Jaws" by \textit{John Williams}. We only consider its bass melody line shown in the first column of Fig. \ref{jaw}. It's very simple, that is, two fixed different pitches jumping back and forth according to the given simple groove\footnote{In music, groove is the sense of propulsive rhythmic "feel" or sense of "swing". In jazz, it can be felt as a persistently repeated pattern. It can be created by the interaction of the music played by a band's rhythm section (e.g. drums, electric bass or double bass, guitar, and keyboards).}.
	
	We can easily find that from the perspective of human perception and experience, this melody is not so flat (even if only considering this melody), maybe we can compare to another melody in the second column of Fig. \ref{jaw}. which has only one single pitch following this groove to get that the previous melody is not so flat, but its rhythm pattern/groove is very flat. So, how to characterize these feelings more rigorous?
	
	\subsubsection{Local/Global Melody Dynamic}
	The first thing we should pay attention to is the Melody Dynamic. As we said above, we need a quantity to help us measure the Melody Dynamic which couples pitch and duration. Before we give the mathematical definition, let us see some examples first.
	
\begin{figure}
		\centering
		\includegraphics{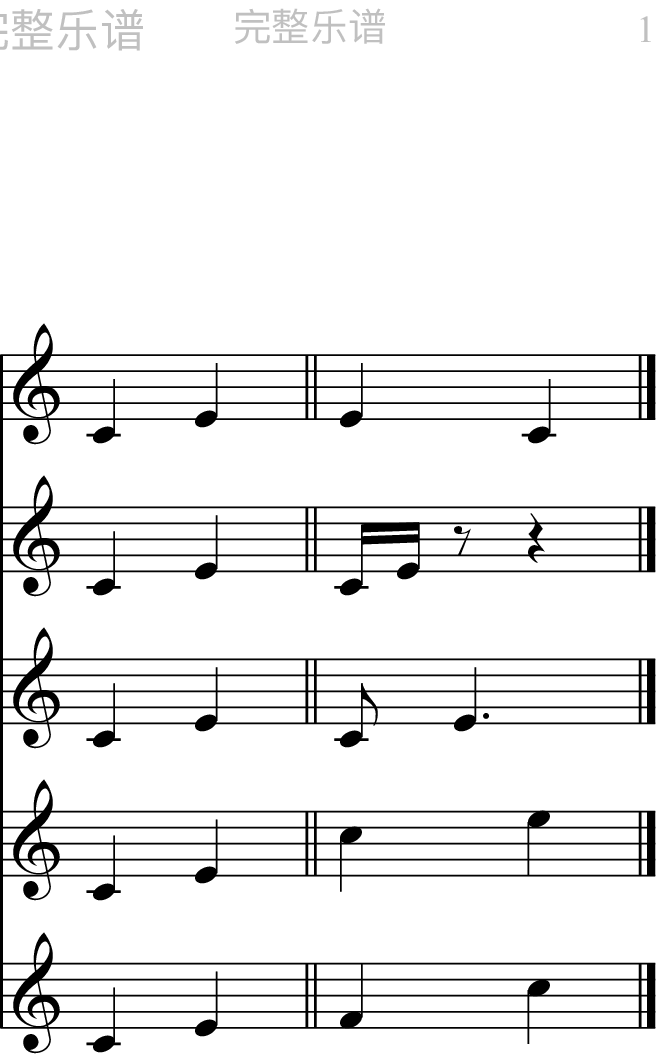}
		\caption{Examples for Local Melody Dynamic}
		\label{lmd_example}
	\end{figure}

	We show some pairs of melody pieces in Fig. \ref{lmd_example}. Both intuitively and experimentally (the experiment result is shown in \ref{choice}, melody pieces in the first column of Fig. \ref{lmd_example} are more flat than the second column, except for the first and fourth row (these are equally flat). 
	
	Next, we formalize the above description and give some rigorous definitions in a mathematical way. Before all that, we assume that for a fixed system, its units of (relative) pitch and duration are fixed\footnote{Note that due to the setting of $pm$, these two units cannot be specified at will. They need to be set according to the actual situation. Here we set a semitone as the pitch unit and a quarter length as a duration unit (experiments in \ref{experiments} will discuss about the rationality of this setting).}.
	
	\hspace*{\fill}
	
	\noindent \textbf{Definition.} Define $pm(N)$ as the \textbf{\textit{Pitch Moment}} of a given note $N=(P,d)$, which is a quantity that couples pitch and duration. Empirically, it should be monotonic about both $P$ and $d$. Here we choose
	\begin{equation}
	pm(N)=pm(P,d)=dP
	\end{equation}
	for simplicity.
	
	\hspace*{\fill}
	
	In some literature of psychological research (such as \cite{temperley2004cognition}), people have studied that for a melody/rhythm, people have a better perception of two adjacent terms which is called pairwise. So here we will introduce a very important concept for the melody dynamic between two successive notes, which we call \textbf{Local Melody Dynamic}.
	
	\hspace*{\fill}
	
	\noindent \textbf{Definition.} Given two successive notes $N_1N_2$, we use
	\begin{equation}
	LMD_{\Phi}(N_1,N_2) = |pm(\hat{N_1})-pm(\hat{N_2})|
	\end{equation}
	to measure their dynamic difference, where $\hat{N_1},\hat{N_2}$ are derived from $\Phi:(N_1,N_2)\rightarrow (\hat{N_1},\hat{N_2})$, where $\Phi$ should allow some basic perceptual assumptions as follows. It's called \textbf{\textit{Local Melody Dynamic(LMD)}} induced by $\Phi$ between $N_1$ and $N_2$.
	
	\noindent \textbf{Perceptual Assumptions of $\Phi$\footnote{The order of these five assumptions is consistent with the column order of Fig. \ref{lmd_example}.}:}\\
	\textbf{$\cdot$ Symmetry of notes:} $LMD_{\Phi}(N_1,N_2)=LMD_{\Phi}(N_2,N_1)$\\
	\textbf{$\cdot$ Decay over time:} $LMD_{\Phi}(N_1',N_2')\leq LMD_{\Phi}(N_1,N_2)$, where $N_1'=(P_1,kd_1), N_2' = (P_2,kd_2), k\geq 1$\\
	\textbf{$\cdot$ Convexity in time:} $LMD_{\Phi}(N_1',N_2')\leq LMD_{\Phi}(N_1,N_2)$, where $N_1'=(P_1,(d_1+d_2)/2), N_2' = (P_2,(d_1+d_2)/2)$\\
	\textbf{$\cdot$ Translation invariance in pitch:} $LMD_{\Phi}(N_1',N_2')= LMD_{\Phi}(N_1,N_2)$, where $N_1'=(P_1+\delta,d_1), N_2' = (P_2+\delta,d_2)$\\
	\textbf{$\cdot$ Monotonicity of pitch interval:} $LMD_{\Phi}(N_1',N_2')\leq LMD_{\Phi}(N_1,N_2)$, where $N_1'=(P_1',d_1), N_2' = (P_2',d_2)$, and $|P_1'-P_2'|\leq|P_1-P_2|$\\
	
	\hspace*{\fill}
	
	In order to illustrate that these assumptions are not contradictory to each other and to facilitate subsequent computations, we directly give a possible $\Phi$ as follows which satisfies all the assumptions,
	\begin{equation}
	\Phi:(N_1,N_2)\rightarrow (\hat{N_1},\hat{N_2})
	\end{equation}
	where
	$$
	\hat{N_1} = (\max\{0,\ P_1-P_2\},\quad \frac{d_1}{d_1+d_2}e^{\frac{1}{d_1}+\frac{1}{d_2}}),
	$$
	$$\hat{N_2} = (\max\{0,\ P_2-P_1\},\quad \frac{d_2}{d_1+d_2}e^{\frac{1}{d_1}+\frac{1}{d_2}}).$$
	
	Finally, we give some quantities of the global dynamic of a whole sentence through the above Local Melody Dynamic.
	
	\hspace*{\fill}
	
	\noindent \textbf{Definition.} Given a monophonic melody $M=N_1N_2...N_n$ without rest, we denote \begin{equation}
	\Delta_{\Phi}M=(LMD_{\Phi}(N_1,N_2),...,LMD_{\Phi}(N_{n-1},N_n))
	\end{equation}
	as $M$'s \textbf{\textit{Local Melody Dynamic Sequence(LMDS)}}. Then we can define
	\begin{equation}
	GMD_{\Phi}(N_1,N_2)=||\Delta_{\Phi}M||
	\end{equation}
	to measure the melody dynamic of the whole sentence, where $||\cdot||$ is a kind of norm. It's called \textbf{\textit{Global Melody Dynamic(GMD)}} induced by $\Phi$ of $M$. Accordingly we can define the \textbf{\textit{normalized GMD(nGMD)}}  as follows,
	\begin{equation}
	nGMD_{\Phi}(N_1,N_2)=\frac{GMD_{\Phi}(N_1,N_2)}{|\Delta_{\Phi}M|}=\frac{||\Delta_{\Phi}M||}{|\Delta_{\Phi}M|}
	\end{equation}
	where $|\Delta_{\Phi}M|=|M|-1$ is the length of Local Melody Dynamic Sequence of $M$.
	
	\hspace*{\fill}
	
	Remark:
	\begin{itemize}
		\item If here we use L1 norm, it’s a bit like total variation\footnote{a mathematical concept from \textit{Real Analysis}} of a given function.
		\item If here we use L$\infty$ norm, it means we use the maximum local melody dynamic to measure the given melody’s global melody dynamic.
		\item Different norms measure different types of dynamic:
		\begin{itemize}
			\item L$\infty$-GMD $\longrightarrow$ irregularity
			\item L1-(n)GMD $\longrightarrow$ time-dependent dynamic
		\end{itemize}
	\end{itemize}
	
	The calculation and visualization demo is shown in \ref{example}.
	
	\subsubsection*{Alternative LMD/GMD}
	Here we also propose an alternative version of LMD and GMD, which is more easy to understand but does not perfectly satisfy the perceptual assumptions required by LMD.
	
	The Alternative LMD/GMD is inspired by \textit{Complex Number}. We consider the pitch and duration of a note as two orthogonal dimensions, that is, the real and imaginary parts of a complex number. Then we can use the angle between two complex numbers to measure the difference between the two notes, in other words, to measure how much effort it takes to move from one note to another (rotation angle). The mathematical definition of Alternative LMD is as follows.
	
	\hspace*{\fill}
	
	\noindent \textbf{Definition.} Given two successive notes $N_1N_2$, we use
	\begin{equation}
	ALMD_{\Phi}(N_1,N_2) = \cos<N_1,N_2>=:\frac{<N_1,N_2>}{|N_1||N_2|}
	\end{equation}
	to measure their dynamic difference alternatively, where $<N_1,N_2> = P_1P_2+d_1d_2, |N_i| = \sqrt{<N_i,N_i>}$. It is called \textbf{\textit{Alternative Local Melody Dynamic}}\footnote{The units of pitch and duration here also need to be considered carefully.}.
	
	\hspace*{\fill}
	
	The definition of \textbf{Alternative Global Melody Dynamic} is similar to before.
	
	\subsubsection{Melody Dynamic Range}
	Above we studied the pairwise melody dynamic in a sentence. But when people listen to music, they can also perceive the difference between the highest and lowest pitch in a sentence or whole music(\cite{temperley2004cognition}), here we call it \textbf{Melody Dynamic Range}.
	
	\hspace*{\fill}
	
	\noindent \textbf{Definition.} Given a monophonic melody $M=K_1K_2...K_n$ where $K$ is a note $N$ or a rest $E$, then we define its \textbf{\textit{Melody Dynamic Range(DR)}} as follows,
	\begin{equation}
	DR(M) = \max_{P\in P(M)}P - \min_{P'\in P(M)}P',
	\end{equation}
	where $P(M)$ is the pitch sequence extracted from $M$ in order.
	
	\hspace*{\fill}
	
	It can be known from the definition and human perception that, the larger the melody dynamic range, the wider the (voice) range of this sentence, and the greater the tension/expression in some senses.
	
	\subsubsection{Rhythm Dynamic}
	As we said before, the melody is composed of notes, and the notes are composed of pitch and duration, then the melody is composed of a pitch sequence and a corresponding duration sequence. Considering only the pitch sequence of melody is missing its musical meaning, that is, it cannot be called music. But considering only the duration sequence of melody has its own musical meaning----it is called rhythm. Above we have delved into the properties of the melody, next we will focus on the rhythm.
	
	In some literatures (\cite{toussaint2012pairwise}\cite{condit2019deconstructing}), people use PVI (Pairwise Variability Index) or nPVI (Normalized Pairwise Variability Index) (defined in \ref{npvi}) to measure the variability of specific time series like rhythm of speech or music. Due to the limitation of these quantities(can not work across style/genre and other things), we design a new framework to analyze the dynamic/variability of the rhythm.
	
	\hspace*{\fill}
	
	\noindent \textbf{Definition.} For a given time series $(d_1,d_2,...,d_n)$, its nPVI is defined as follows,
	\begin{equation}
	\label{npvi}
	\textbf{nPVI}=\frac{100}{n-1}\sum_{i=1}^{n-1}|\frac{d_i-d_{i+1}}{(d_i+d_{i+1})/2}|.
	\end{equation}
	
	\hspace*{\fill}
	
	First, we expanded the definition and framework of PVI and nPVI like LMD and GMD, to study the rhythm in more detail.
	
	\hspace*{\fill}
	
	\noindent \textbf{Definition.} Given a monophonic melody $M=K_1K_2...K_n$ where $K$ is a note $N$ or a rest $E$, then we use $R(M) = d_1d_2...d_n$ to denote its rhythm, which means $R(M)$ only take the duration of notes and rests in $M$. 
	
	Let $\Delta^{(1)} R(M)=(d'_1,...,d'_{n-1})$, where $d'_i=|\frac{d_{i+1}-d_i}{(d_{i+1}+d_i)/2}|$, we call it \textbf{\textit{PV Sequence}}. Then we have
	\begin{equation}
	\textbf{nPVI}=100\frac{||\Delta^{(1)} R(M)||_1}{|\Delta^{(1)} R(M)|}.
	\end{equation}
	
\begin{figure}
		\centering
		\includegraphics[scale=0.7]{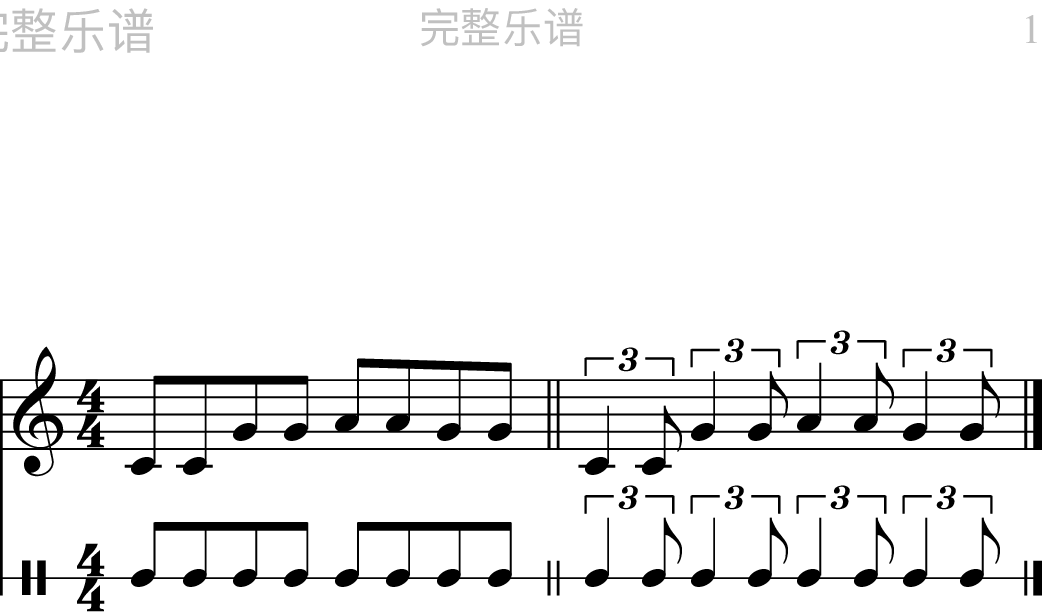}
		\caption{The first measure's nPVI is 0, and the second measure's nPVI is 66.66.}
		\label{rd_example}
	\end{figure}

	But even the refined/expanded PVI has limitations. Let's see an example in Fig. \ref{rd_example}.
	
	Here we don't discuss the definition of style/genre too much. We only assert that,\\
	\textbf{If we only changed the groove (such as making a song swing), the rhythm dynamic of the melody itself will not be changed. In other words, the Rhythm Dynamic of a given melody should be a property that is style/genre-agnostic and inherent in itself.} 
	As for how to distinguish the two melodies in Fig. \ref{rd_example}, the previously defined melody dynamics are enough (the melody dynamics of the two melody are 42.64 and 182.57 respectively).
	
	So we need to define a style/genre-agnostic Rhythm Dynamic. In other words, the Rhythm Dynamics of the two melodies in the above example should be the same. Next we give a mathematical definition of Rhythm Dynamic.
	
	\hspace*{\fill}
	
	\noindent \textbf{Definition.} (All notations are the same as \textbf{\textit{PV Series}}.) Let $\Delta^{(2)} R(M)= \Delta^{(1)}\Delta^{(1)} R(M) = \Delta^{(1)}(d'_1,...,d'_{n-1})=(d_1'',...,d_{n-2}'')$, where $d''_i=|\frac{d'_{i+1}-d'_i}{(d'_{i+1}+d'_i)/2}|$\footnote{If the denominator in one of the above functions is equal to $0$, the value of the function is taken as $0$.},  it requires $n>2$,\footnote{This is consistent with human perception. Because it ’s hard to say what the inherent rhythmic dynamic of the two successive notes is, because they do not move forward.} we call this quantity as \textbf{\textit{Rhythm Dynamic Series}}, and call its norm 
	\begin{equation}
	RD(M) = ||\Delta^{(2)}R(M)||
	\end{equation} 
	as \textbf{\textit{Rhythm Dynamic}} (no Global), 
	\begin{equation}
	nRD(M) = \frac{RD(M)}{|M|-2}
	\end{equation}
	as \textbf{\textit{Normalized Rhythm Dynamic}}.
	
	\hspace*{\fill}
	
	We calculate the rhythm dynamics of the two melodies in Fig. \ref{rd_example} by the above definition, after that we can get that they are both $0$. 
	
	
	\subsection{Examples}
	\label{example}
	The second column of Fig. \ref{cherbourg} shows our analysis of a complete song in the first column, we plot sentence-wise MCG, GMD and RD of it. We can see from these figures that when the chorus part (B) is reached, although the RD is very small (which means the rhythm is relatively flat), both MCG and GMD are bigger than those in verse (A\& C). This shows that although it is relatively flat compared to verse from the perspective of rhythm, the melody dynamic in chorus are still bigger than verse. In line with human perception, it illustrates the rationality of our definition of Melody Dynamic and Rhythm Dynamic.
	
\begin{figure}
		\subfigure{
			\begin{minipage}[]{0.7\linewidth}
				\includegraphics[height=4.5cm, width=5.7cm]{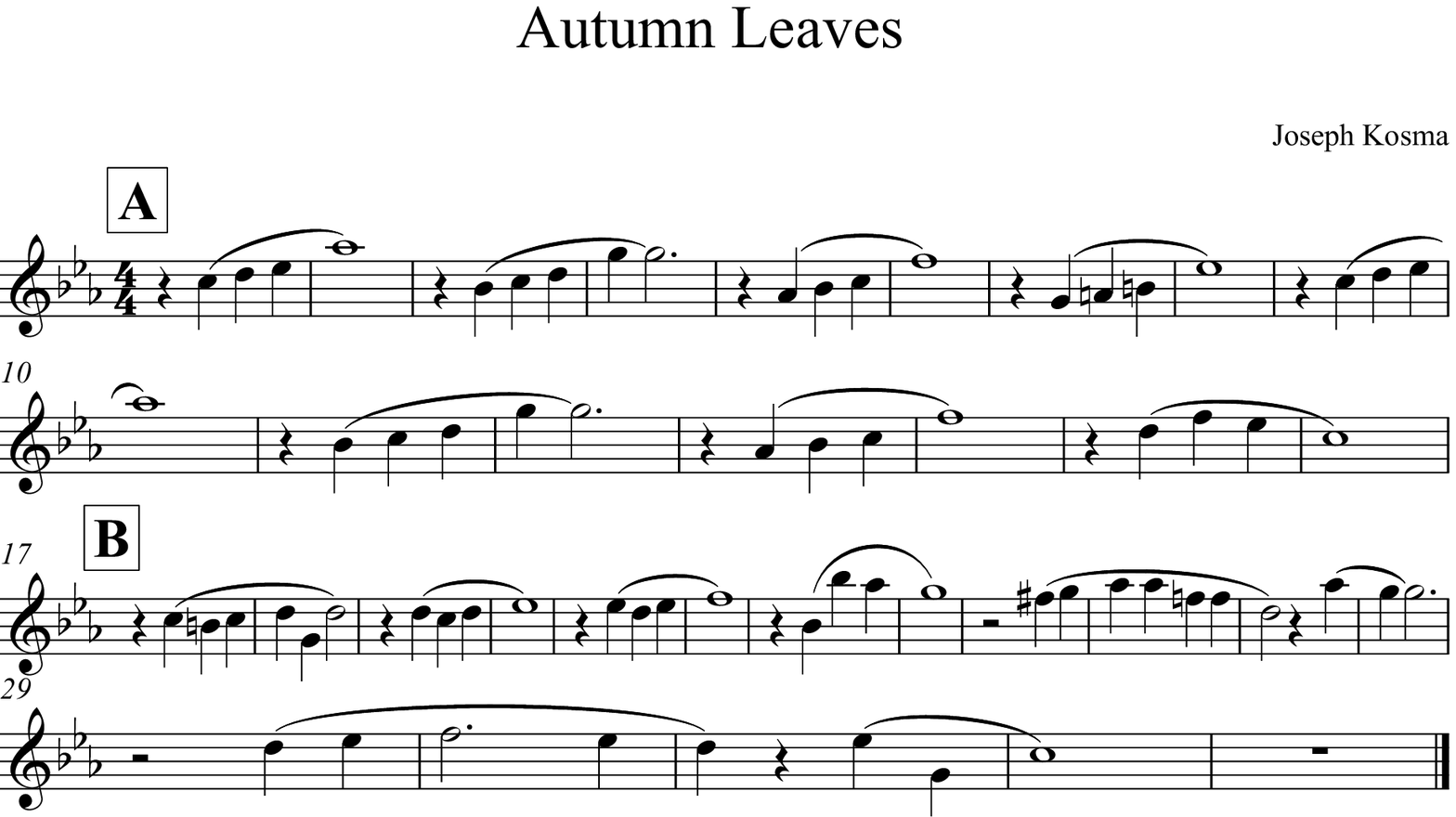}
			\end{minipage}
			\begin{minipage}[]{0.2\linewidth}
				\includegraphics[scale=0.15]{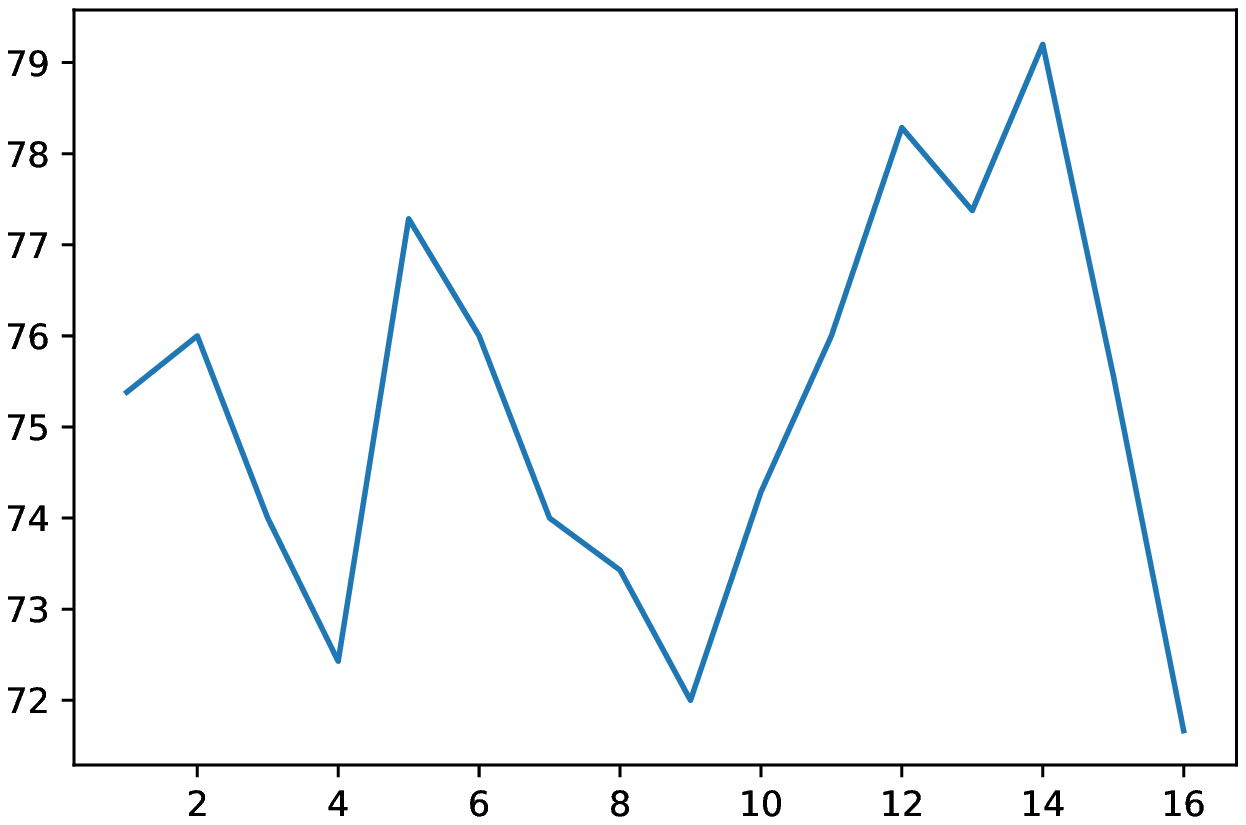}
				\includegraphics[scale=0.15]{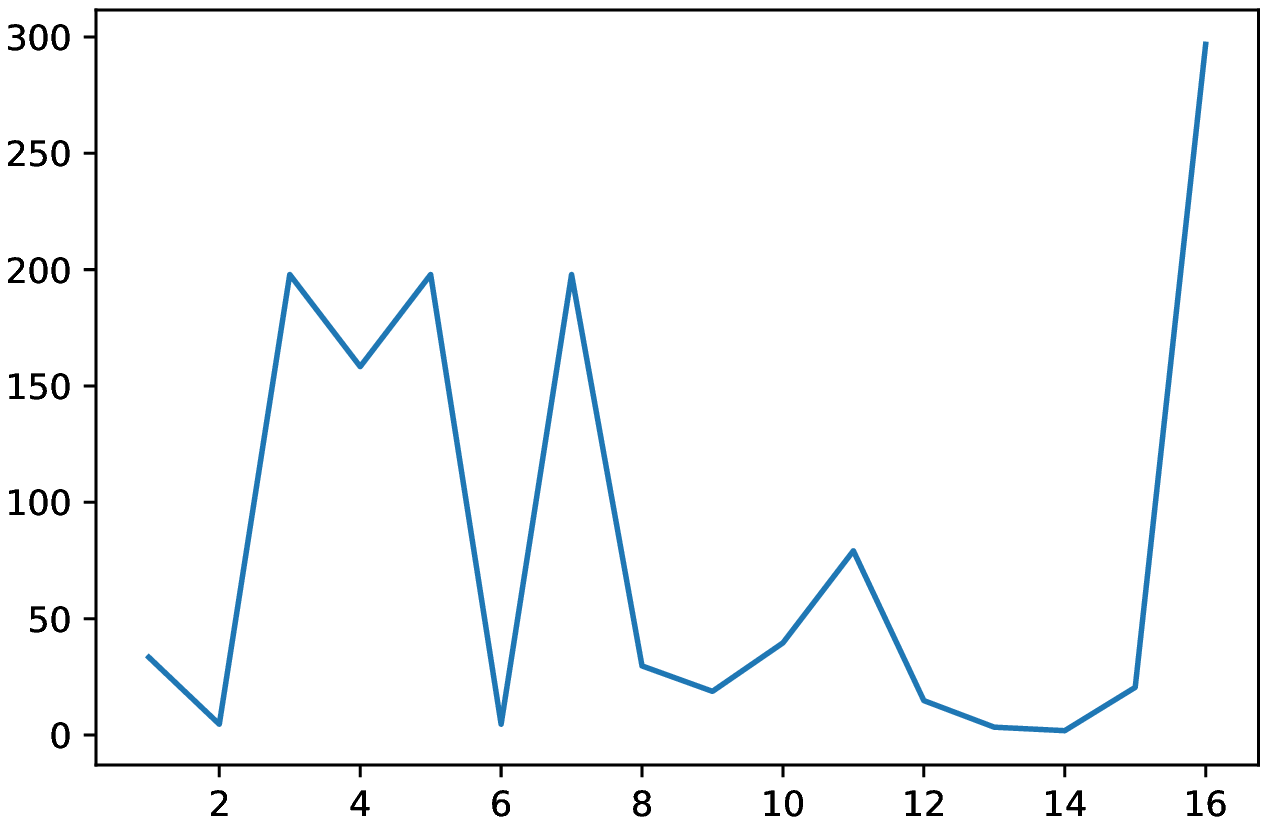}
				\includegraphics[scale=0.15]{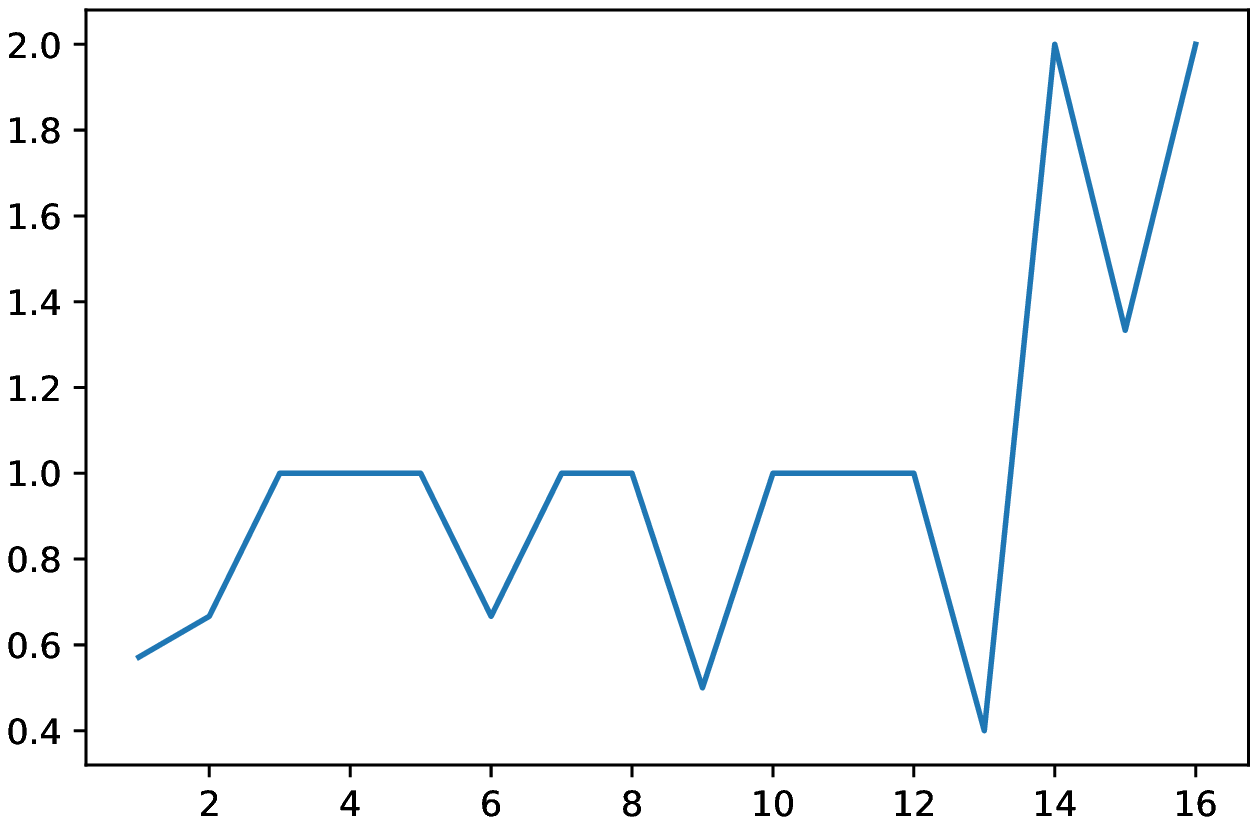}
			\end{minipage}%
		}%
		\caption{Autumn Leaves}
		\label{al}
	\end{figure}

\begin{figure}
		\subfigure{
			\begin{minipage}[]{0.7\linewidth}
				\includegraphics[height=4.5cm, width=5.7cm]{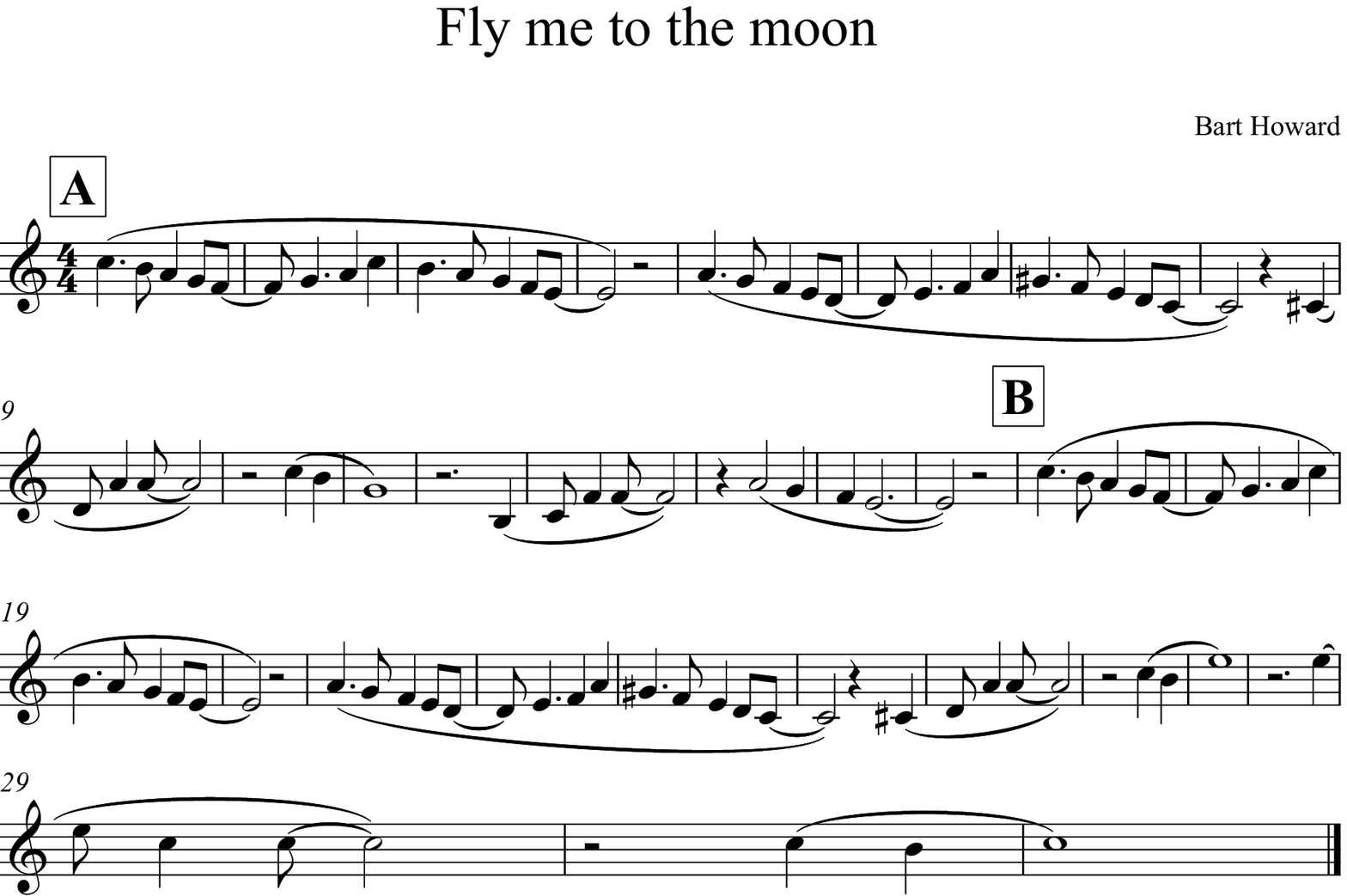}
			\end{minipage}
			\begin{minipage}[]{0.2\linewidth}
				\includegraphics[scale=0.15]{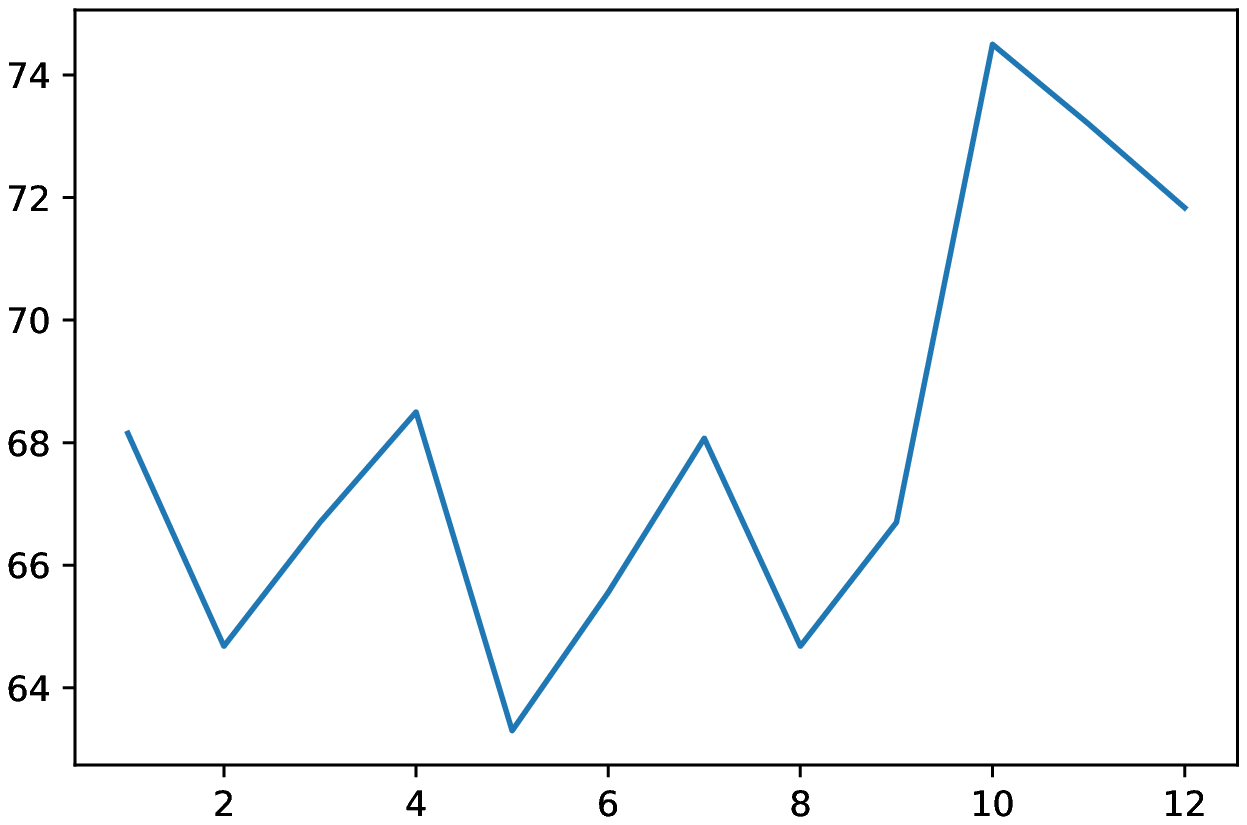}
				\includegraphics[scale=0.15]{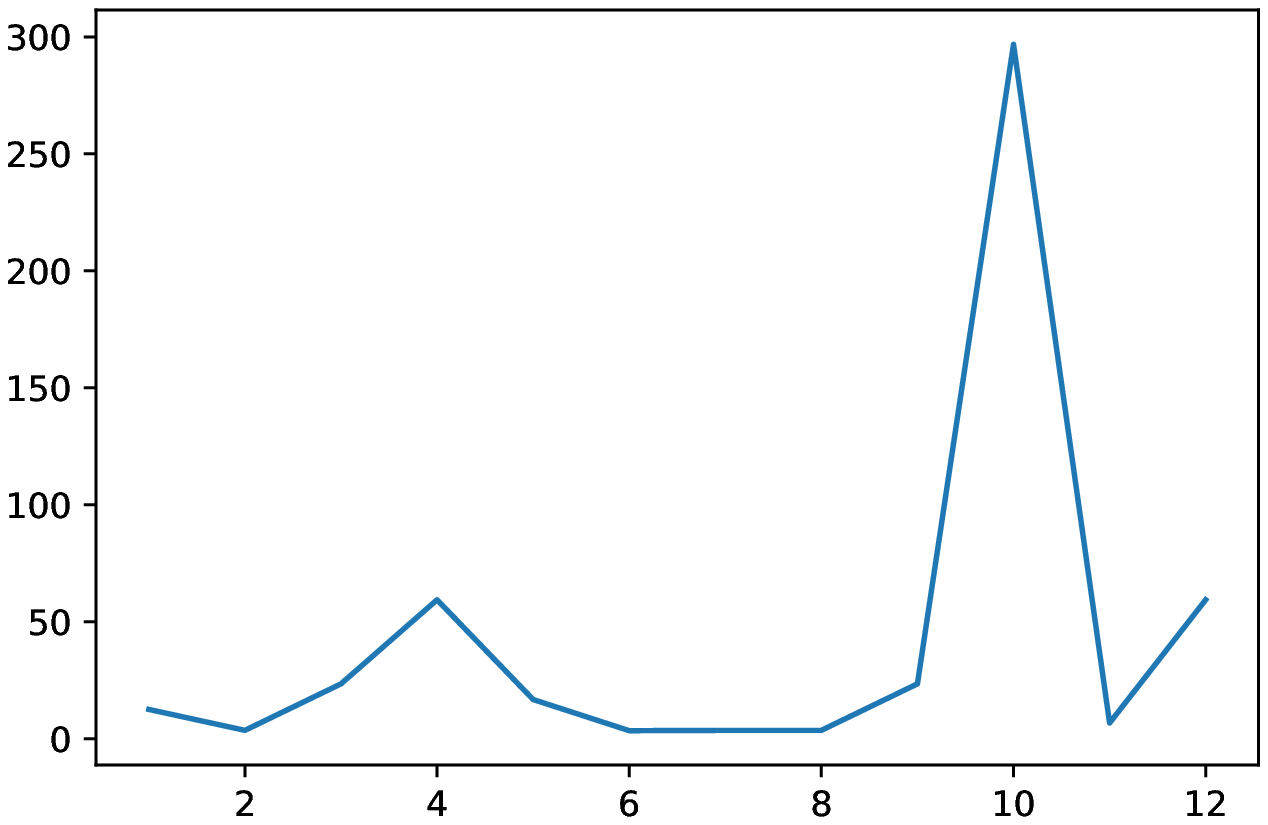}
				\includegraphics[scale=0.15]{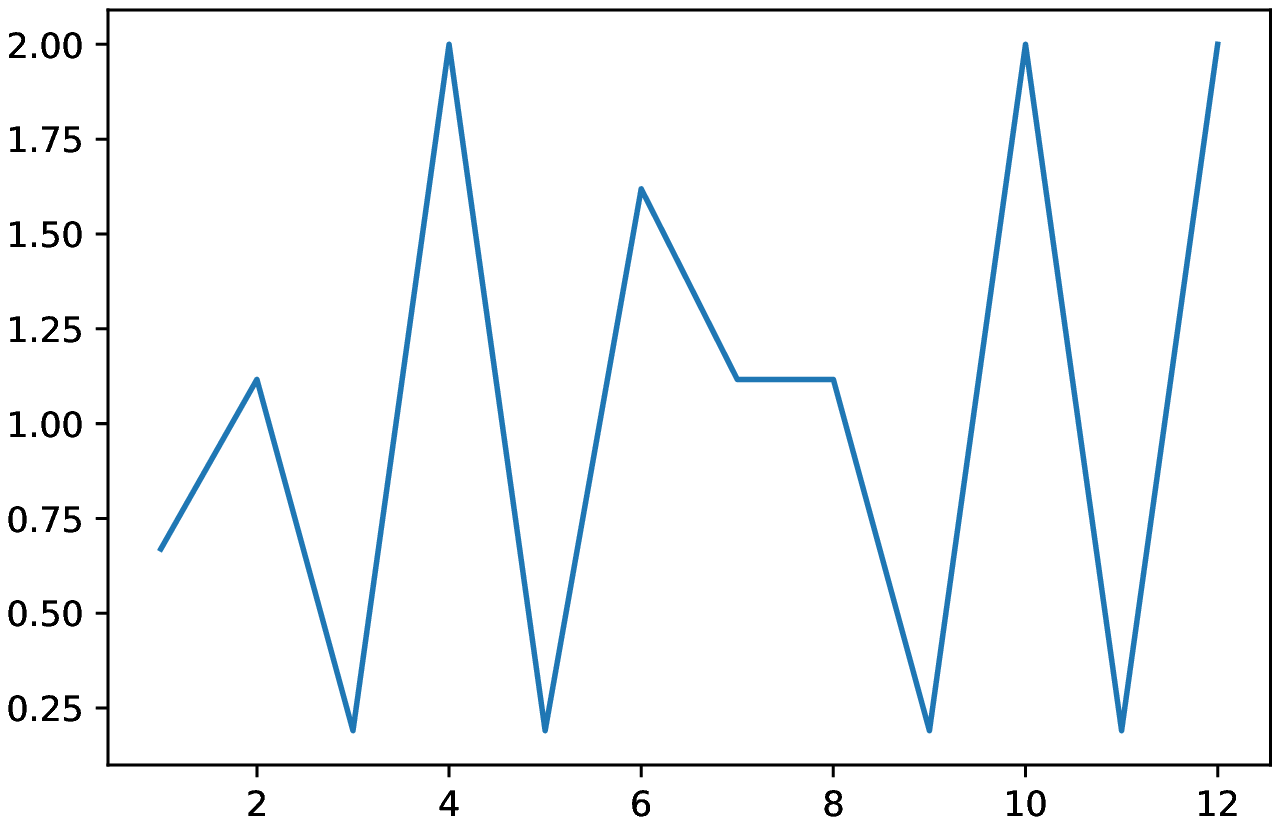}
			\end{minipage}%
		}%
		\caption{Fly Me to the Moon}
		\label{fl}
	\end{figure}

\begin{figure}
		\subfigure{
			\begin{minipage}[]{0.7\linewidth}
				\includegraphics[height=4.5cm, width=5.7cm]{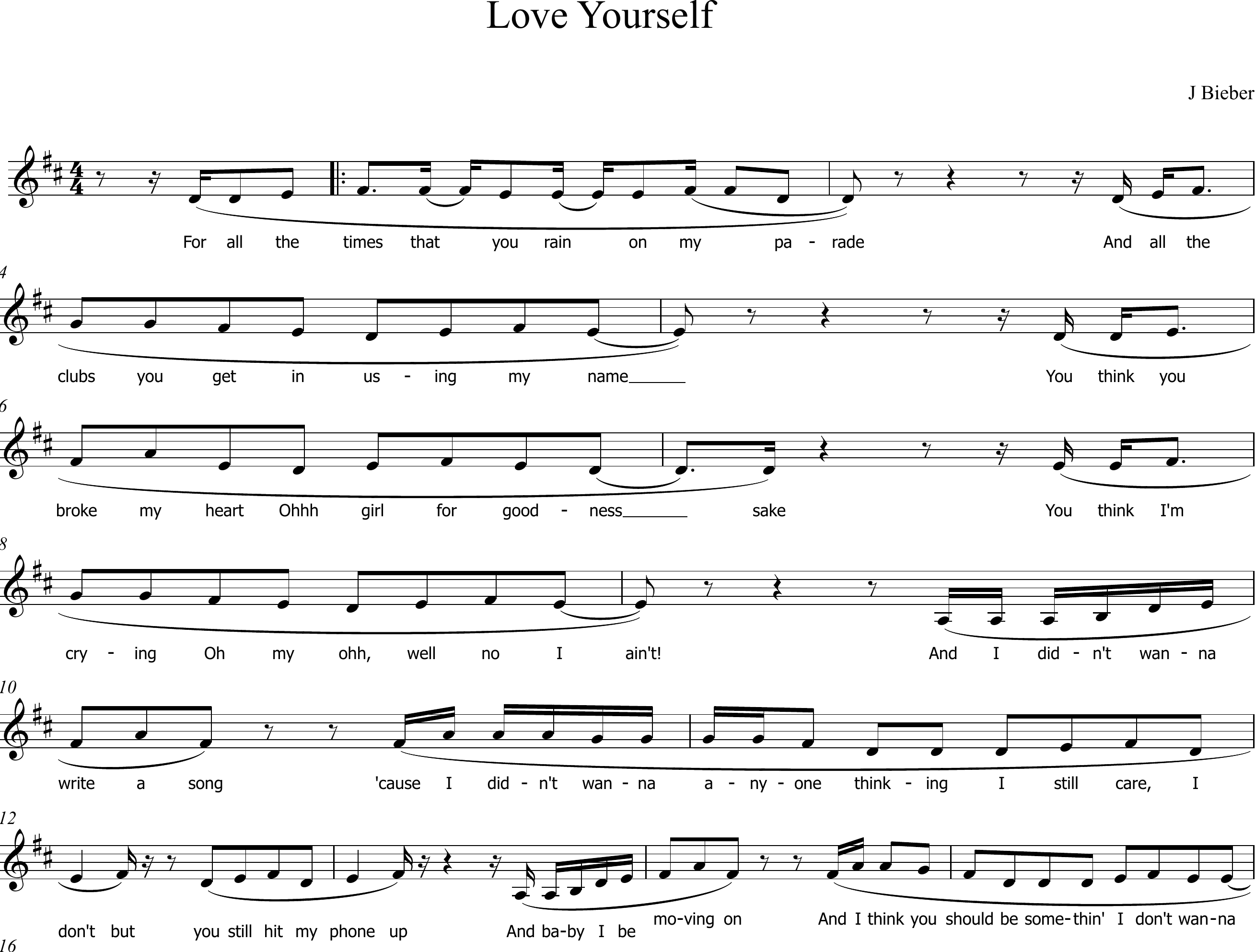}
			\end{minipage}
			\begin{minipage}[]{0.2\linewidth}
				\includegraphics[scale=0.15]{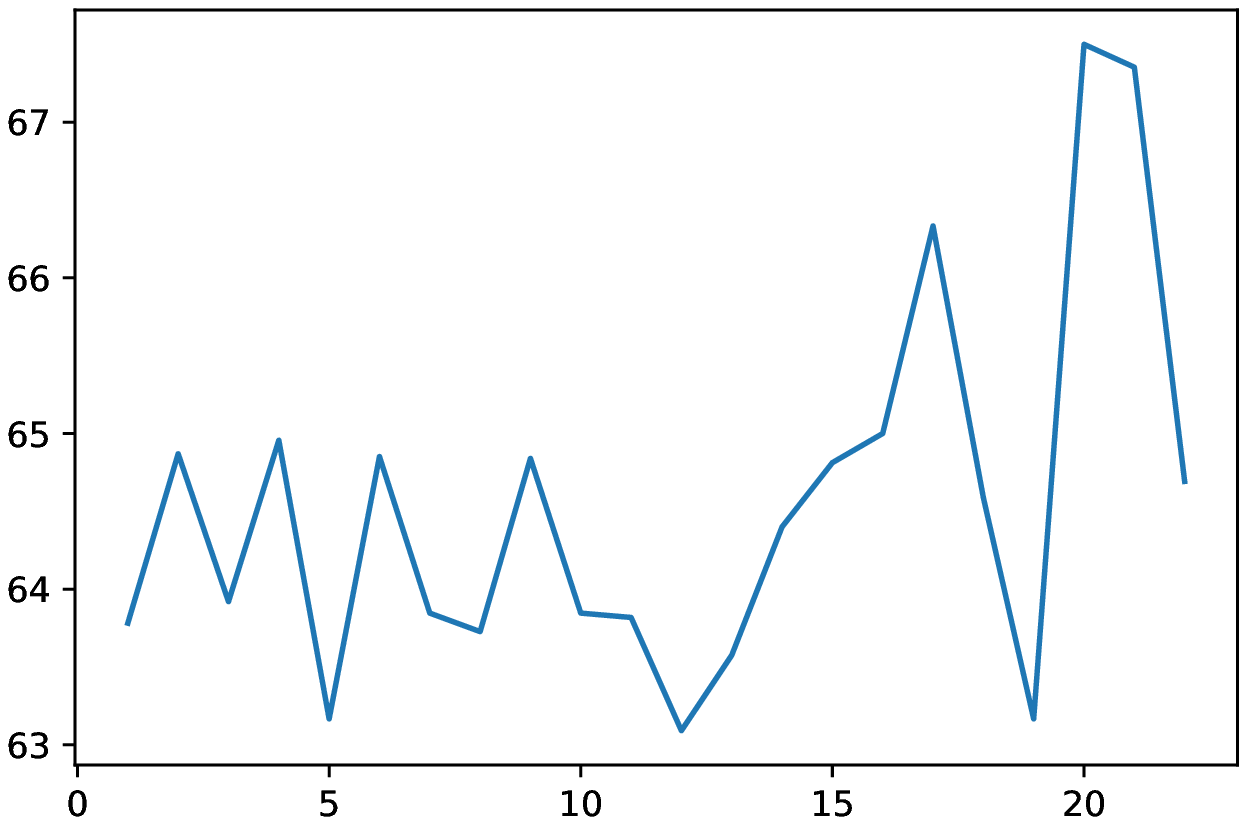}
				\includegraphics[scale=0.15]{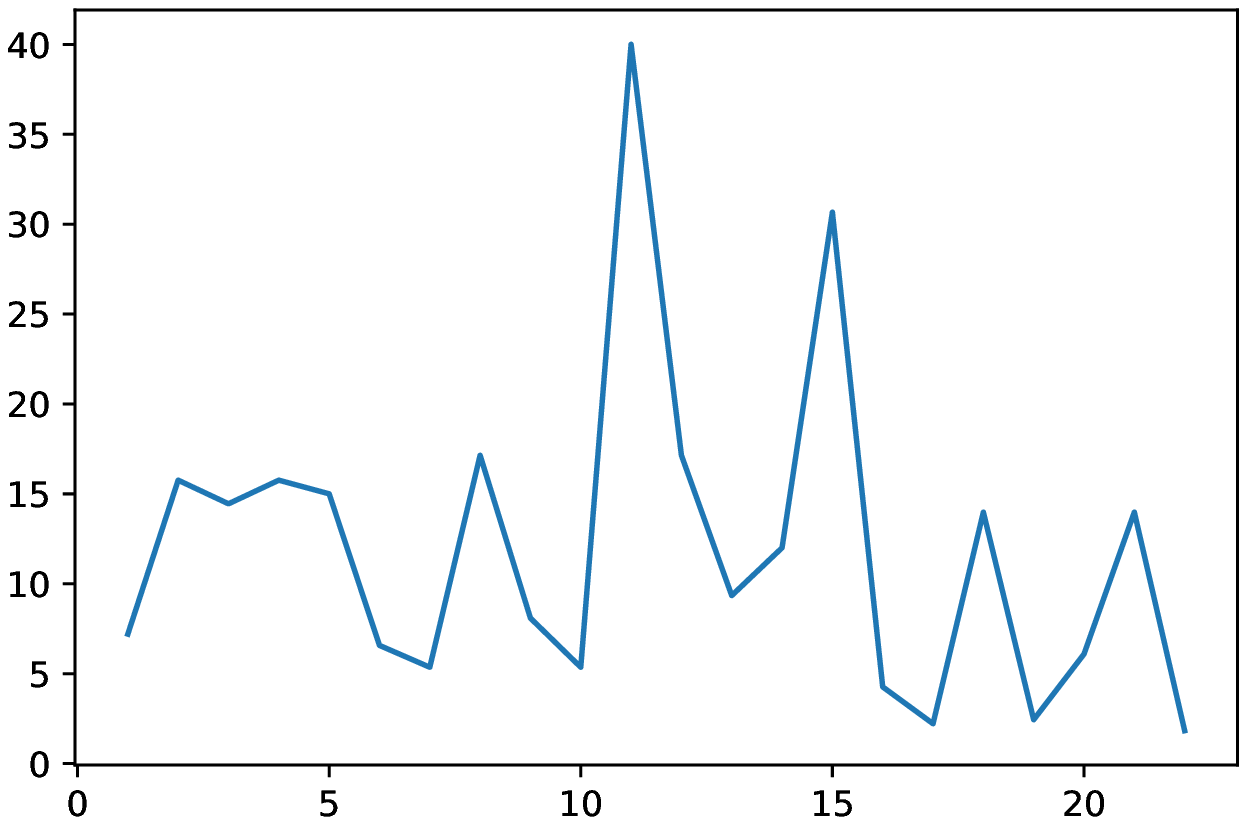}
				\includegraphics[scale=0.15]{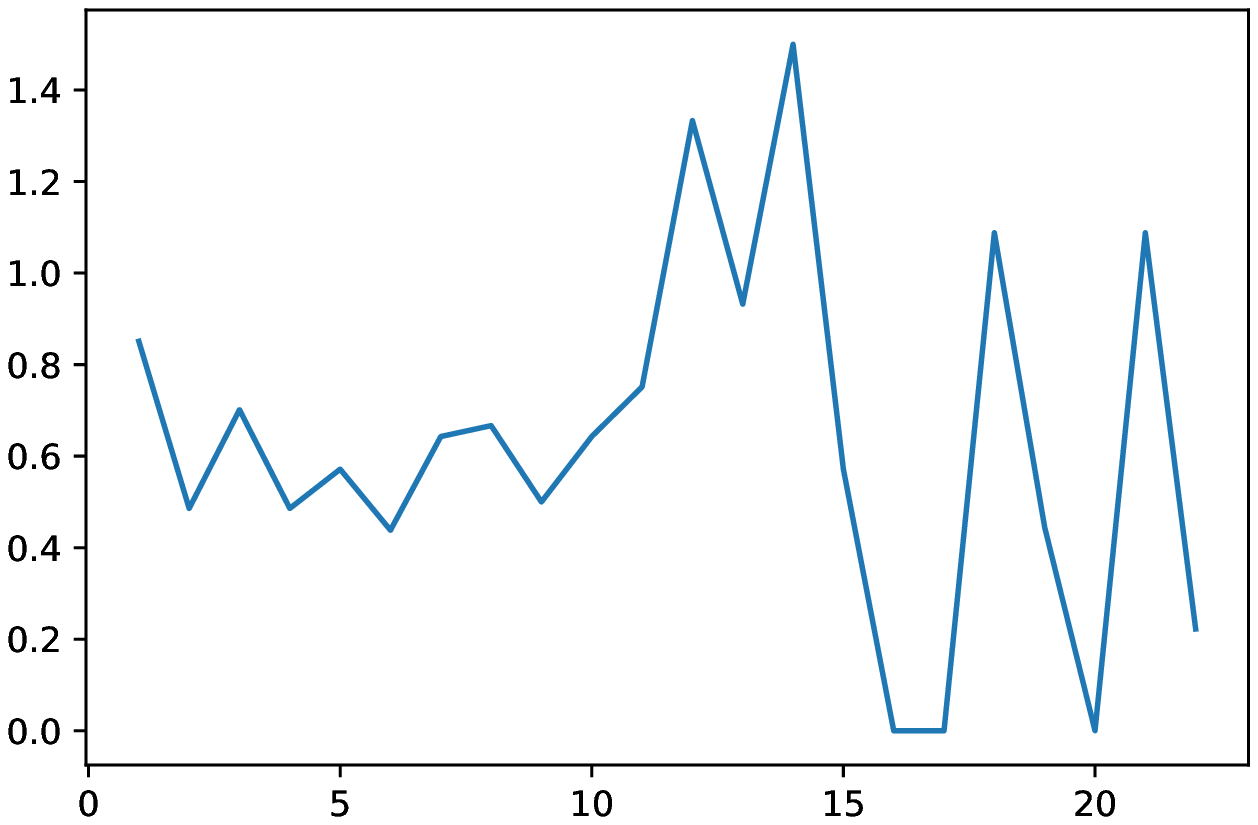}
			\end{minipage}%
		}%
		\caption{Love Yourself (part)}
		\label{ly}
	\end{figure}

	Here we show three more songs and their analyses in Fig. \ref{al}, \ref{fl} and \ref{ly}. In these 3 song we choose normalized L$\infty$-GMD, because the sentences in these songs are not so neat, normalized GMD would be more appropriate.
	
	\label{proof}
	
	\section{Experiments}
	\label{experiments}
	\subsection{Choice of the pitch and time duration units}
	\label{choice}
	Here we briefly describe the experimental process and results. We have 10 participants (2 pros, 3 enthusiasts and 5 ordinary people) in this experiment. For each participant, we let them listen to 10 music from the real-world music dataset mentioned in \ref{realworld}, and plot 9 versions of pictures which are similar to the second column of Fig. \ref{cherbourg}. The difference between these nine versions is the choice of the pitch and time duration units. Each participant selects a set of pictures (MCG, GMD and RD) that he thinks best represent his perception of the song, and the final score is the cumulative number of votes for each version. Before the experiment, each participant was explained in detail what these three quantities (MCG, GMD and RD) should represent. Details are in the table \ref{unit}.
	
	From this table, we can conclude that (1, 1) is the most consistent with human perception among the nine choices.
	
	\begin{table}[]
		\centering
		\begin{tabular}{c|c|c|c}
			\hline
			Version  & Pitch unit & Duration unit & Score\\
			\hline
			1  & 0.5 & 0.5 & 3\\
			\hline
			2  & 0.5 & 1 & 5\\
			\hline
			3  & 0.5 & 2 & 4\\
			\hline
			4  & 1 & 0.5 & 7\\
			\hline
			5  & 1 & 1 & 56\\
			\hline
			6  & 1 & 2 & 8\\
			\hline
			7  & 2 & 0.5 & 2\\
			\hline
			8  & 2 & 1 & 3\\
			\hline
			9  & 2 & 2 & 2\\
			\hline
		\end{tabular}
		\caption{Pitch unit/semitone, Duration unit/quarterlength}
		\label{unit}
	\end{table}

	\subsection{Extract these features in real-world music dataset}
	\label{realworld}
	We have two real-world music dataset, one is Hooktheorytab (lead sheet) dataset\footnote{\href{https://github.com/wayne391/lead-sheet-dataset}{https://github.com/wayne391/lead-sheet-dataset}}, the other is a C-pop dataset which is made by ourselves. We selected some data on the two data sets for experiments, and all the experimental data are in our github repo\footnote{\href{https://github.com/water45wzh/exploring_melody}{https://github.com/water45wzh/exploring\_melody}}. All the code and useful data are shown in this repo.

	\section{Conclusion and Future Work}
	The importance of studying the inherent properties of melody is self-evident, whether for objective evaluation of a melody or creating melodies by machine based on rules or deep learning. Due to the low-dimensionality of highly structured music data, it's difficult to use deep learning or other dimensionality reduction method to do feature engineering to extract inherent interpretable properties for it. In this paper, we focus on the inherent properties of the monophonic melody of songs. First, we define the melody mathematically. Then we define some coarse-grained quantities based on melody composition technique and human perception (such as Melodic Center of Gravity, Local/Global Melody Dynamic, etc.) to study the inherent properties of melody, we noticed that in a sense, some of these quantities are style/genre-agnostic, and can work across different temperaments or tonality/atonality. Finally, we take some real-world songs as examples to study the meaning of these quantities. Although we have not proposed criteria for evaluating whether a melody is good or bad, just propose some novel interpretable quantities to characterize some of the inherent properties of the melody. We believe, and try to apply these features to promote research on the objective evaluation or automatic generation of melody based on rules or deep learning in the future.

	\bibliography{smc2020bib}
	
\end{document}